\documentclass[ aip, amsmath, amssymb, reprint]{revtex4-1}

\usepackage{graphicx}% Include figure files
\usepackage{dcolumn}% Align table columns on decimal point
\usepackage{bm}% bold math

\usepackage[utf8]{inputenc}
\usepackage[T1]{fontenc}
\usepackage{mathptmx}
\usepackage{subfigure}

\usepackage[table, dvipsnames]{xcolor}

\usepackage{hyperref}

\hypersetup{
    colorlinks=true,
    linkcolor=blue,
    citecolor=blue,
    filecolor=magenta,      
    urlcolor=blue,
}

\newcommand{\comment}[1]{}
\begin{document}

\preprint{AIP/123-QED}

\title[]{Photoelectron spectra of early $3d$-transition metal dioxide cluster anions from $GW$ calculations}

\author{Meisam  Rezaei and Serdar \"{O}\u{g}\"{u}t}
\affiliation{Department of Physics, University of Illinois at Chicago, Chicago, IL 60607, USA}
\email{ogut@uic.edu.}
\date{\today}

\begin{abstract}
Photoelectron spectra of early $3d-$transition metal dioxide anions, ScO$_{2}^-$, TiO$_{2}^-$, VO$_{2}^-$, CrO$_{2}^-$, MnO$_{2}^-$, are calculated using semilocal and hybrid density functional theory (DFT) and many-body perturbation theory within the $GW$ approximation using one-shot perturbative and eigenvalue self-consistent formalisms. Different levels of theory are compared with each other and with available photoelectron spectra. We show that one-shot $GW$ with a PBE0 starting point ($G_0W_0$@PBE0) consistently provides very good agreement for all experimentally measured binding energies (within 0.1-0.2 eV or less), which we attribute to the success of PBE0 in mitigating self-interaction error and providing good quasiparticle wave functions, which renders a first-order perturbative $GW$ correction effective. One-shot $GW$ calculations with semilocal exchange in the DFT starting point ({\em e.g.} $G_0W_0$@PBE) do poorly in predicting electron removal energies by underbinding orbitals with typical errors near 1.5 eV. Higher amounts of exact exchange ({\em e.g.} 50\%) in the DFT starting point of one-shot $GW$ do not provide very good agreement with experiment by overbinding orbitals with typical errors near 0.5 eV. While not as accurate as $G_0W_0$@PBE0, the $G$-only eigenvalue self-consistent $GW$ scheme with $W$ fixed to the PBE level ($G_nW_0$@PBE) provides a reasonably predictive level of theory (typical errors near 0.3 eV) to describe photoelectron spectra of these $3d-$transition metal dioxide anions. Adding eigenvalue self-consistency also in $W$ ($G_nW_n$@PBE), on the other hand, worsens the agreement with experiment overall. Our findings on the performance of various $GW$ methods are discussed in the context of our previous studies on other transition metal oxide molecular systems.
\end{abstract}

\maketitle

%%%%%%%%%%%%%%%%%%%%%%%%%%%%%%%%%%% Introduction %%%%%%%%%%%%%%%%%%%%%%%%%%%%%%%%%%

\section{\label{sec:Intro}Introduction}

The last decade has witnessed a growing number of computational studies that have benchmarked Green's function methods, such as the $GW$ approximation\cite{Reining2018} and the Bethe-Salpeter equation, for excited state properties of bulk and molecular systems.\cite{Rinke2005,Bruneval2006,Tiago2006,Shishkin2007,Fuchs2007,Rostgaard2010,Ma2010,Blase2011-a,Blase2011-b,Faber2011,Qian2011,Ke2011,Ren2012,Marom2012,Baumeier2012,Setten2013,Bruneval2013,Pham2013,Caruso2013,Atalla2013,Klimes2014,Faber2014,Koval2014,Boulanger2014,Korbel2014,Wang2015,Hirose2015,Bruneval2015,Jacquemin2015,Setten2015,Wilhelm2016,Kaplan2016,Caruso2016,Knight2016,Hung2016,Hung2017a,Hung2017b,Maggio2017,Govoni2018,Grumet2018,Shi2018,Byun2019,Gao2019} A large majority of these studies have focused on the performance of various flavors of the $GW$ approximation in predicting the electron removal energies in $sp$-bonded molecules and clusters by comparing their predictions with accurate quantum chemistry calculations and experimental photoelectron spectroscopy data. Modeling excited states of systems containing transition metal elements in general, and transition metal oxides in particular, within the $GW$ theory have faced additional theoretical and computational challenges,\cite{Golze2019} as enhanced electron correlations inherent in these systems and their propensity to having open-shell electronic configurations necessitate the use of more sophisticated approaches beyond simple perturbative implementations on top of density functional theory (DFT) with semi-local exchange-correlation functionals. Furthermore, convergence issues with respect to basis set size and other implementation parameters present significant computational bottlenecks for maintaining a desirable level of accuracy comparable to what can be achieved for $sp$-bonded systems that is typically $\sim 0.1$ eV. Accordingly, there have been much fewer studies on the performance of the $GW$ approximation for systems that contain transition metal elements. Motivated by this observation, here we continue with our recent benchmark studies\cite{Hung2017a,Hung2017b,Shi2018,Byun2019} by focusing on the electronic structure of negatively charged $3d$-transition metal dioxide clusters TMO$_2^-$, for TM = Sc, Ti, V, Cr, and Mn.

Photoelectron spectra of early $3d$ TMO$_2^-$ clusters have been available up to photon energies of $5-6.5$ eV since the pioneering studies of Wang and collaborators from more than two decades ago.\cite{Wu1997,Wu1998-a,Wu1998-b,Gutsev2000-b,Gutsev2001,Zhai2007,Kim2014} Their structural and electronic properties have been investigated\cite{Knight1996,Chertihin1997,Walsh1999,Zhou1999,Gonzales2000,Vyboishchikov2000,Gutsev2000-a,Dong2002,Grein2007,Li2008,Uzunova2008,Gong2009,Liu2009,Lee2009,Qu2010,Marom2011,Taylor2012,Kim2013,Hendrickx2014,Xu2015} in various computational studies using methods based on DFT and quantum chemistry. While these TMO$_2^-$ molecules are isostructural with little changes in the bond angles and lengths upon changing the TM element, their frontier molecular orbitals display a wide range of spatial localization properties and have significantly varying amounts of TM $3d$ and O $2p$ contents. Therefore, their electronic structures provide a diverse set of challenges to DFT within the generalized Kohn-Sham scheme and to many-body perturbation theory within the $GW$ approximation, due to the need to mitigate self-interaction error (SIE) and the importance of a suitable DFT starting point or a form of self-consistency in their $GW$ treatment. In this work, we model the photoelectron spectra (PES) of early TMO$_2^-$ clusters using DFT with semi-local and hybrid exchange-correlation functionals (based on shifted eigenvalue spectra) and within the $GW$ approximation using one-shot perturbative schemes with different DFT starting points as well as eigenvalue self-consistent formalisms. 

This article is organized as follows. We begin with an overview of the computational methods and parameters used in our calculations in Sec. \ref{sec:Computational detail}. This is followed in Sec. \ref{sec:results} with a brief discussion of the structural and vibrational properties of the five TMO$_2^-$ clusters considered in this study and detailed analyses of their PES computed with DFT and $GW$ methods in comparison to each other and to experimental data. The overall trends in the performance of shifted DFT and variants of the $GW$ approximation are discussed in Sec. \ref{sec:Trend}. Finally, we summarize our findings and analyses in Sec. \ref{sec:summary}. 

%%%%%%%%%%%%%%%%%%%%%%%%%%%%%%%%%%% computational details %%%%%%%%%%%%%%%%%%%%%%%%%%%%%%%%%%

\section{\label{sec:Computational detail}Computational details}

DFT computations to find the optimized structures were carried out with the NWCHEM code,~\cite{Valiev2010} Version 6.6, using PBE exchange-correlation functional~\cite{Perdew1996a} and aug-cc-pVTZ basis sets. The photoelectron spectra were simulated by broadening the eigenvalue spectra with a Gaussian distribution function of 0.1 eV smearing width, without taking into account the photoionization cross sections. For DFT spectra obtained with PBE and PBE0\cite{Perdew1996b} functionals, the Kohn-Sham eigenvalues were shifted to align the first peak with the vertical ionization potential (IP) of the anion, which is computed as the total energy difference between the anionic and neutral clusters at the fixed anionic geometry.
  
$GW$ calculations were carried out using the MOLGW~\cite{Bruneval2016} software within both the one-shot (perturbative) $G_0W_0$ scheme and the eigenvalue self-consistent $GW$ (ev$GW$) scheme with two types of self-consistency, $G_nW_0$ and $G_nW_n$, which update the eigenvalues only in $G$ and in both $G$ and $W$, respectively. In order to investigate the starting point dependency of the $GW$ approximation, we used global hybrid functionals
\begin{eqnarray}
{E_{\text{xc}}^{\text{PBE}\alpha}}=\alpha E_{\text{x}}^{\text{HF}} + (1-\alpha)E_{\text{x}}^{\text{PBE}} + E_{\text{c}}^{\text{PBE}},
\end{eqnarray}
where $E_{\text{x}}^{\text{HF}}$, $E_{\text{x}}^{\text{PBE}}$, $E_{\text{c}}^{\text{PBE}}$ and $\alpha$ are the Fock exact exchange, PBE exchange, PBE correlation energies, and the amount of exact exchange, respectively. For starting point dependency of the $G_0W_0$ scheme, we tested three different values for $\alpha$: 0 (PBE starting point), 0.25 (PBE0 starting point) and 0.50. In these full-frequency $G_0W_0$ calculations, we solved the nonlinear quasiparticle equation for each state graphically using the secant (quasi-Newton) method. The numerical parameter $\eta$ used to broaden the self-energy poles was chosen as $\eta=0.001$ Hartree, and the self-energy was evaluated on a frequency grid of spacing $\Delta\omega=0.001$ Hartree. As discussed in detail in Ref. \onlinecite{Byun2019}, full frequency $G_0W_0$ method leads to complicated self-energy pole structures, typically at nonfrontier orbitals, which results in multiple solutions of the quasiparticle equation and makes determining the correct and accurate quasiparticle energies difficult. This multisolution issue of $G_0W_0$ is particularly prevalent in transition metal oxide systems with a PBE starting point. As such, we checked the accuracy of our predictions for the quasiparticle energies by (i) also using the spectral-function method, which locates the quasiparticle peak with the highest weight in the spectral function, (ii) looking at the trends as a function of the basis set size and the amount of exact exchange, and (iii) by varying the parameters $\eta$ and $\Delta\omega$, as recommended in Ref. \onlinecite{Byun2019}. 

For calculations performed with the $G_nW_0$ and $G_nW_n$ self-consistency methods, we used the iterative scheme employed in MOLGW, described in Ref. \onlinecite{Byun2019}, where the quasiparticle renormalization factor $Z$ is set to $Z=1$. In these calculations, we used a larger broadening parameter $\eta=0.01$ Hartree for the self-energy poles (and accordingly, a larger frequency grid spacing of $\Delta\omega=0.01$ Hartree) in order to avoid the oscillations typically observed in the quasiparticle energies of nonfrontier orbitals as a function of the iteration index. With these parameters, self-consistency to 0.01 eV was achieved within 3-9 iterations depending on the starting point, basis set size, and the particular orbital considered. In this paper, we only present results from $G_nW_0$ and $G_nW_n$ calculations with a PBE starting point, as we observed that $G_nW_0$ and $G_nW_n$ schemes with PBE0 (or larger values of exact exchange) lead to worse agreement with experiment. The results for $G_nW_0$ with PBE0 and hybrid functional with $\alpha=0.5$ starting points are presented for completeness in the supplementary material. 

Our $GW$ calculations were performed using aug-cc-pVTZ and aug-ccpVQZ basis sets and extrapolated to the complete basis set (CBS) limit with the following function~\cite{Hatting2012}
\begin{eqnarray}
E=a + \frac{b}{N_{\rm BF}},
\end{eqnarray}
where $E$ is a quasiparticle energy, $a$ and $b$ are the fitting parameters, and $N_{\rm BF}$ is the number of basis functions. We tested the accuracy of these fits by performing the $G_0W_0$ calculations with various starting points for a closed-shell (ScO$_2^-$) and open-shell (TiO$_2^-$) molecule with the aug-cc-pV5Z basis sets. We observed that including aug-cc-pV5Z basis sets only had a negligible effect (few tens of meVs) on the CBS limit. We note, however, that extrapolation to the CBS limit is a must for accurate predictions of quasiparticle energies in transition metal oxide molecular systems, as the CBS limit is typically lower than the value obtained with aug-cc-pVQZ basis sets by appreciable amounts ranging from approximately 0.2 eV to 0.6 eV depending on the molecule and the particular orbital considered. These findings are consistent with trends observed in Ref. \onlinecite{Byun2019}. Finally, the Coulomb interaction terms were evaluated using the resolution-of-identity (RI) approximation.\cite{Hill2008,Ren2012} We checked the accuracy of the RI approximation at both the DFT and $G_0W_0$ levels, and the differences were found to be negligible, typically in the $1-10$ meV range.

\section{\label{sec:results}Results}

The optimized structural parameters of the molecular anions considered in this study are shown in Table~\ref{tab:structure}. All molecules have the $C_{2v}$ symmetry. Our results for the TM-O bond lengths and the O-TM-O bond angles (obtained with PBE functional) are in excellent agreement with the results of Gutsev {\em et al.}\cite{Gutsev2000-a} Also shown in Table~\ref{tab:structure} are the harmonic vibrational frequencies, which again agree very well with the results of Gutsev {\em et al.}, and the ground state symmetries and spin multiplicities of the molecules. ScO$_2^-$ is the only molecule with a closed-shell ground state; all others are open-shell ranging from doublet for TiO$_2^-$ to quintet ground state for MnO$_2^-$. Next, we discuss the electronic structures and the photoelectron spectra obtained at various levels of DFT and $GW$ theory for each of the five transition metal dioxide anions. For this discussion, we place the molecules in the $yz$ plane and choose $z$ as the $C_2$ symmetry axis. We use the convention in which the $b_1$ orbital is antisymmetric with respect to reflection in the $yz$ plane and $b_2$ is symmetric. We employ Mulliken population analysis to quantify the $3d$ characters of the molecular orbitals. 

%%%%%%%%%%%%%%%%%%%%%%%%%%% TableI: Bond length, bond angle and vibrational frequencies  %%%%%%%%%%%%%%%%%%%%%%%%%%%%%%%%%%%%

\begin{table*}
\begingroup
\setlength{\tabcolsep}{10pt} % Default value: 6pt
\renewcommand{\arraystretch}{2} % Default value: 1
\caption{Computed TM-O bond lengths, O-TM-O bond angles, vibrational frequencies, and ground state symmetries for the TMO$_2^-$ molecules considered in this study.}
\label{tab:structure}
\begin{tabular*}{1.00\textwidth}{ @{\extracolsep{\fill}} l c c c c c c }
\hline \hline

%\cline{6-7}
                                &   Bond length (\AA)  & Bond angle ($^\circ$)  & $\omega_1 (a_1)$ & $\omega_2 (a_1)$ & $\omega_3 (b_2)$ &  Ground state  \\  \hline
ScO$_2^{-}$                     &   1.81    &   124.1   &   184 &   769 &   671 &   $^1A_1$ \\ \hline

TiO$_2^{-}$                     &   1.682   &   111.4   &   321 &   924 &   888 &   $^2A_1$ \\ \hline

VO$_2^{-}$                      &   1.653   &   117.9   &   293 &   908 &   907 &   $^3B_{1}$ \\ \hline

CrO$_2^{-}$                     &   1.649   &   134.0   &   242 &   867 &   921 &   $^4B_{1}$   \\ \hline

MnO$_2^{-}$                     &   1.662   &   126.0   &   250 &   879 &   835 &   $^5B_{2}$   \\ \hline

\hline \hline
\end{tabular*}
\endgroup
\end{table*}

%%%%%%%%%%%%%%%%%%%%%%%%%%%%%%%%%%% ScO2-1 %%%%%%%%%%%%%%%%%%%%%%%%%%%%%%%%%%

\subsection{\label{sec:ScO2-}  ScO$_\textbf{2}^-$}

\begin{figure}
\includegraphics[scale=0.5]{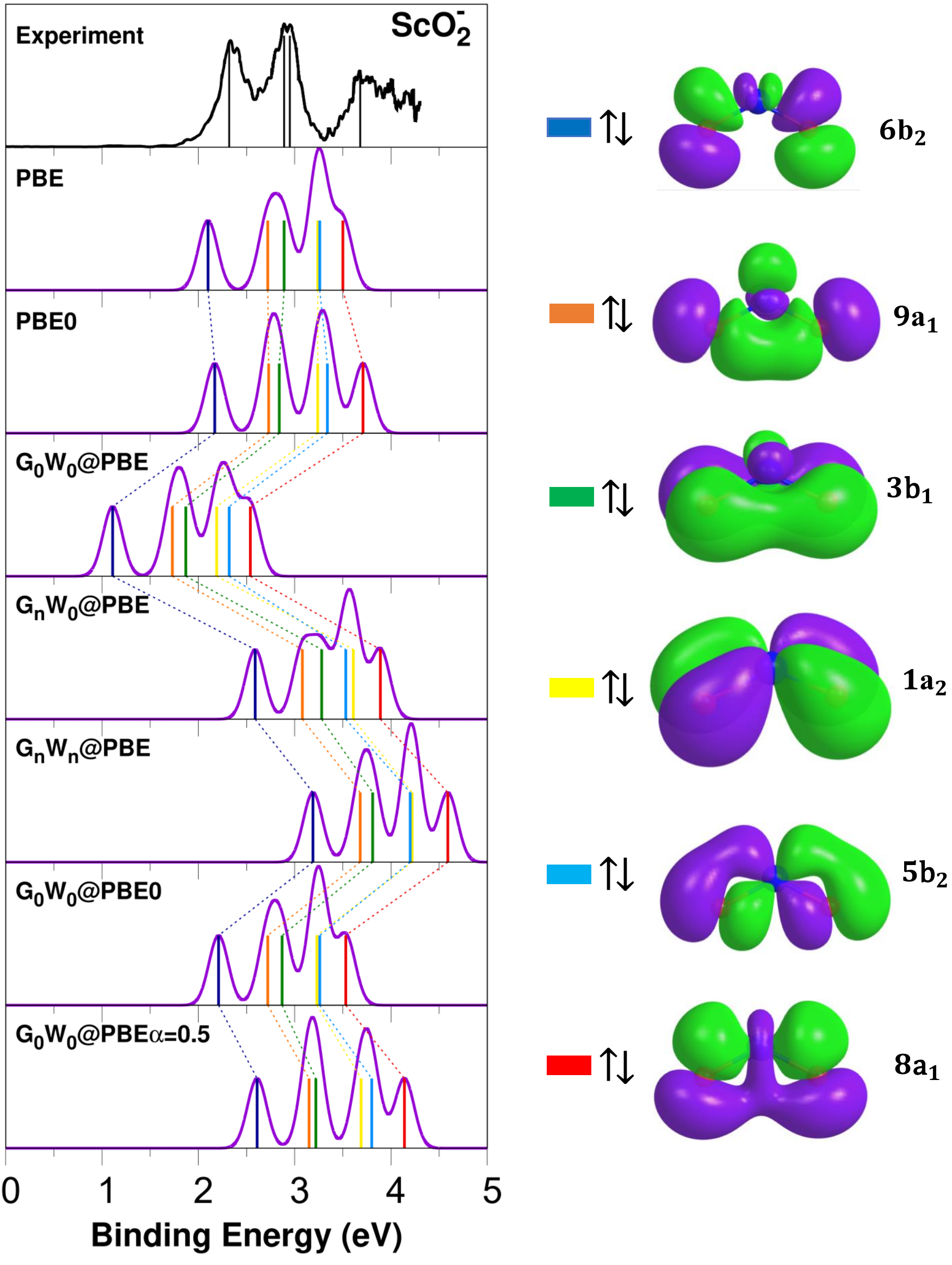}
\caption{\label{fig:ScO2} Experimental photoelectron spectrum of ScO$_{2}^-$ (Ref.~\onlinecite{Wu1998-a}) along with spectra computed with shifted PBE, shifted PBE0, $G_0W_0@\text{PBE}$, $G_nW_0@\text{PBE}$, $G_nW_n@\text{PBE}$, $G_0W_0@\text{PBE0}$ and $G_0W_0$@PBE{\scriptsize$\alpha=0.5$}. Contour plots for some of the occupied molecular orbitals are shown on the right, with matching
color codes displayed in the spectra.}
\end{figure}

Figure \ref{fig:ScO2} shows the experimental photoelectron spectrum\cite{Wu1998-a} of ScO$_2^-$ along with the spectra calculated at various levels of theory. The experimental spectrum obtained with a 266 nm (4.66 eV) laser consists of three main features, starting with a peak centered at 2.32 eV, followed by a more intense and broad peak near 2.9 eV (attributed in Ref. \onlinecite{Wu1998-a} to two overlapping features at 2.89 and 2.95 eV) and a higher energy broad peak centered at 3.68 eV. With PBE (as well as PBE0) functional, we find the ground state of ScO$_2^-$ to be a singlet with a valence configuration $(8a_1)^2(5b_2)^2(1a_2)^2(3b_1)^2(9a_1)^2(6b_2)^2$. We find the ground state of the neutral molecule to be a doublet with the electron removed from the $6b_2$ orbital. The vertical IPs are calculated as 2.10 and 2.17 eV with PBE and PBE0 functionals, respectively, which compare reasonably well with the experimental value of 2.32 eV. Both the shifted PBE and PBE0 spectra agree quite well with experimental data for the first three peaks, while they underestimate the fourth peak by 0.44 eV, which corresponds to the removal of $1a_2$ (or $5b_2$) electron with both functionals. In transition metal oxide molecules, shifted PBE spectra do not typically lead to good agreement with experimental data for localized orbitals or orbitals with considerable $3d$ character, as shown for the case of copper oxide molecular anions in Ref. \onlinecite{Shi2018}. We attribute the apparent success of the shifted spectra of ScO$_2^-$ computed with PBE for the first three peaks to the observation that the three highest occupied orbitals $(3b_1,9a_1,6b_2)$ are delocalized and primarily due to O $2p_x$, $2p_y$, and $2p_z$ states, respectively, with negligible ($<6\%$) Sc $3d$ character. The orbitals with the largest ($\sim20$\%) $3d$ content for this molecule are the $1a_2$ (with $d_{xy}$ character) and $5b_2$ (with $d_{xz}$ character) orbitals, for which the agreement with experimental data is not as satisfactory.\cite{Wu1998-a} Both the shifted PBE and shifted PBE0 spectra have the same mean absolute error (MAE) of 0.22 eV.

Among the variants of the $GW$ approximation, the predictions from the one-shot $GW$ with a PBE starting point ($G_0W_0$@PBE) are particularly poor. All quasiparticle energies are underestimated significantly (by more than $\sim$1.1 eV) compared to experiment, leading to a large MAE of 1.24 eV at this level of theory. Predictions from $G_0W_0$ calculations with hybrid functional starting points are much better, as shown in Fig. \ref{fig:ScO2}. In fact, both $G_0W_0$@PBE0 and $G_0W_0$@PBE{\scriptsize$\alpha=0.5$} levels of theory have the lowest MAE of 0.21 eV averaged over the four experimentally measured peaks. It is important, however, to note that not all peaks are predicted equally well at these two levels of theory: $G_0W_0$@PBE0 predictions for the first three peaks are quite accurate (within $\sim0.15$ eV), while the fourth peak is underestimated by 0.45 eV, very similar to the case of the shifted PBE0. $G_0W_0$@PBE{\scriptsize$\alpha=0.5$} predictions, on the other hand, are not as good for the first three peaks (off by $\sim0.3$ eV), while the fourth peak is predicted within 0.01 eV. These findings correlate with the amount of $3d$ character of the relevant orbitals: Typically, orbitals with larger $3d$ content require a larger amount of exact exchange in the $G_0W_0$ starting point, as also observed for the case of  copper oxide molecular anions.\cite{Shi2018} Another level of theory that leads to good agreement with experimental data is $G_nW_0$@PBE, which has been argued to be a practical $GW$ scheme for the electronic structure of transition metal oxide molecular systems as a good compromise between computational efficiency and accuracy.\cite{Byun2019} The MAE for $G_nW_0$@PBE is 0.24 eV. Applying eigenvalue self-consistency also in $W$ ($G_nW_n$), on the other hand, deteriorates the agreement with experimental data, and the MAE for this level of theory is 0.76 eV.

%%%%%%%%%%%%%%%%%%%%%%%%%%%%%%%%%%% TiO2-1 %%%%%%%%%%%%%%%%%%%%%%%%%%%%%%%%%%

\subsection{\label{sec:TiO2-} TiO$_\textbf{2}^-$}

\begin{figure}
\includegraphics[scale=0.5]{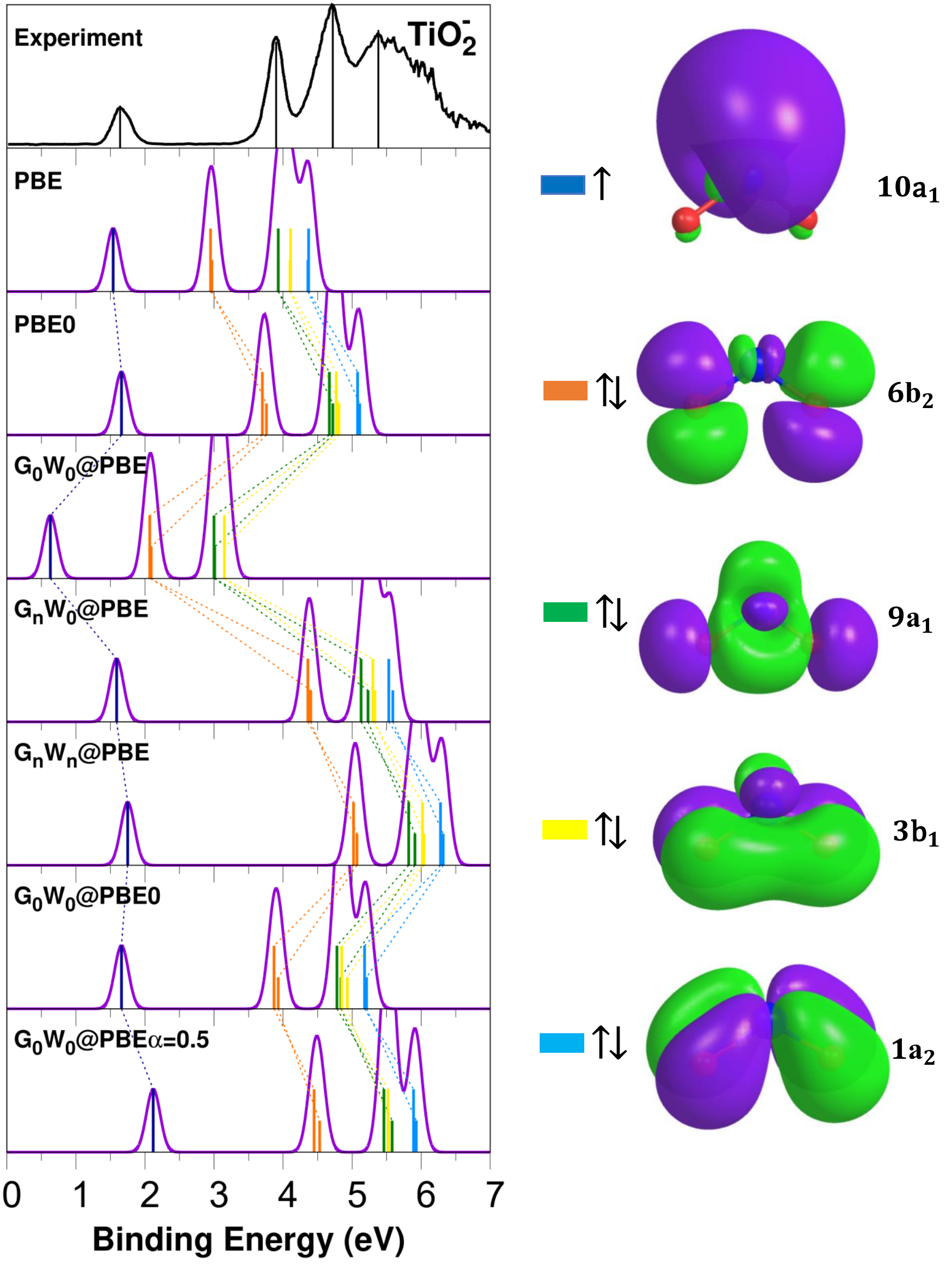}
\caption{\label{fig:TiO2} Experimental photoelectron spectrum of TiO$_{2}^-$ (Ref.~\onlinecite{Zhai2007}) along with spectra computed with shifted PBE, shifted PBE0, $G_0W_0@\text{PBE}$, $G_nW_0@\text{PBE}$, $G_nW_n@\text{PBE}$, $G_0W_0@\text{PBE0}$ and $G_0W_0$@PBE{\scriptsize$\alpha=0.5$}. Contour plots for some of the occupied molecular orbitals, in both majority ($\uparrow$) and minority ($\downarrow$) spin channels, are shown on the right, with matching color codes displayed in the spectra. For doubly occupied orbitals, the calculated energy displayed with the higher (lower) height refers to the orbital in the majority (minority) spin channel.}
\end{figure}

Figure \ref{fig:TiO2} shows the experimental photoelectron spectrum of TiO$_2^-$ and the spectra calculated with shifted DFT and various $GW$ flavors. The most recent experimental spectrum obtained with a 193 nm (6.42 eV) laser consists of four main peaks at 1.60, 3.90, 4.72, and 5.38 eV.\cite{Zhai2007} With both PBE and PBE0 functionals, we find the ground state of TiO$_{2}^-$ to be a doublet ($^2A_1$) with a valence configuration $(1a_2)^2(3b_1)^2(9a_1)^2(6b_2)^2(10a_1)^1$, and the ground state of the neutral molecule is a singlet with the electron removed from the very delocalized $10a_1$ orbital with large Ti $4s$ content and small Ti $p_z$ and $3d_{x^2-z^2}$ admixtures. As in ScO$_2^-$, the $6b_2$ state is entirely due to O $2p_z$ states, but it is considerably more localized in TiO$_2^-$. The higher binding energy (BE) orbitals $9a_1$ and $3b_1$ are also more localized in TiO$_2^-$, and they have larger $3d$ content ($\sim$15\%) compared to ScO$_2^-$. 

The vertical IPs calculated with PBE and PBE0 functionals are 1.54 and 1.66 eV, bracketing the experimental value of 1.60 eV. In the following comparisons, we compare the experimental values for the higher BE peaks with the average of the calculated spin-up and spin-down eigenvalues of a given doubly occupied orbital. While the vertical IPs calculated with both PBE and PBE0 are in very good agreement with experiment, shifted PBE clearly fails to provide satisfactory predictions when higher BE peaks are taken into account, as shown in Fig.~\ref{fig:TiO2}. Shifted PBE0, on the other hand, performs quite well for the first three peaks and underestimates the energy of the fourth. This observation is another example of the well-known failure of semi-local functionals, such as PBE, as they underbind localized orbitals due to SIE, while the addition of a fraction of exact exchange globally, as in PBE0, or in range-separated form helps to mitigate this error.\cite{Shi2018} The overall MAE of shifted PBE and PBE0 are found to be 0.77 and 0.21 eV, respectively. These observations are in good agreement with the results of Marom {\em et al.}\cite{Marom2011} who reported similar findings.

Among the variants of the $GW$ approximation, $G_0W_0$@PBE has the worst performance with an MAE of 1.68 eV. Similar to ScO$_2^-$, the best overall agreement with experiment is achieved with the $G_0W_0$@PBE0 level of theory (with an MAE of 0.16 eV), while adding more exact exchange worsens the agreement with experiment, as the MAE of $G_0W_0$@PBE{\scriptsize$\alpha=0.5$} is 0.51 eV, significantly larger than that of ScO$_2^-$. The effect of eigenvalue self-consistency in predicting quasiparticle energies of TiO$_2^-$ is similar to that of ScO$_2^-$. $G_nW_0$ provides fairly good agreement with experiment (MAE of 0.25 eV), while iterating eigenvalues in $W$ worsens the agreement by overbinding all orbitals other than the highest occupied molecular orbital (HOMO), leading to an MAE of 0.77 eV. 
% We can perhaps discuss the disagreement with Noa's G0W0@PBE0 results. We can decide later.

%%%%%%%%%%%%%%%%%%%%%%%%%%%%%%%%%%%  VO2-1 %%%%%%%%%%%%%%%%%%%%%%%%%%%%%%%%%%

\subsection{\label{sec:VO2-} VO$_\textbf{2}^-$}

Previous studies on the electronic structure of VO$_2^-$ have revealed many possible low energy configurations as candidates for its ground state.\cite{Chertihin1997,Gutsev2000-a,Vyboishchikov2000,Uzunova2008,Kim2014,Hendrickx2014} Based on an earlier study on the electronic structure of the neutral molecule,\cite{Knight1996} Wu and Wang interpreted their experimental photoelectron spectrum of VO$_2^-$ by assuming that the ground state of the anion is a singlet of $^1A_1$ symmetry.\cite{Wu1998-b} Later computational studies showed that the $^1A_1$ state is higher in energy than the triplet states $^3B_1$ and $^3A_1$ by $0.3-0.5$ eV. However, depending on the electronic structure method and computational details, such as the basis set size and the choice of the exchange-correlation functional in DFT treatments, different studies have found the ground state to be either $^3B_1$ or $^3A_1$, typically by a very small energy difference. For example, the most recent DFT calculations of Kim {\em et al.}\cite{Kim2014} and RCCSD(T) calculations of Hendrickx and Tran\cite{Hendrickx2014} have found $^3B_1$ as the ground state by small differences of 0.02 to 0.07 eV relative to $^3A_1$. 

In our DFT calculations with PBE functional, we find the ground state of VO$_2^-$ to be the $^3B_1$ triplet with the valence configuration $(3b_1)^2(9a_1)^2(6b_2)^2(10a_1)^1(4b_1)^1$. The ground state of the neutral molecule is found to be a doublet with the electron removed from the $4b_1$ orbital, which has a large ($\sim$70\%) $3d_{xz}$ character and is significantly more localized than the $10a_1$ orbital. As a result, with the addition of a fraction of exact exchange, which reduces SIE, the $4b_1$ orbital moves down in energy much more than the $10a_1$ orbital. Accordingly in our calculations with the PBE0 functional, while the ground state of VO$_2^-$ is still found to be the $^3B_1$ triplet, the HOMO is the $10a_1$ orbital. The downward shifts in the Kohn-Sham eigenvalues of the $4b_1$ and $10a_1$ orbitals in going from PBE to PBE0 are 1.83 and 0.90 eV, respectively, which leads to the observed reordering of the orbitals. We note that our PBE calculations find the $^1A_1$ singlet state to be 0.4 eV higher in energy, in agreement with previous DFT and quantum chemistry calculations. Below, we only discuss the spectra obtained for the triplet configuration. As we show in the supplementary material, none of the theoretical spectra obtained for the singlet configuration resembles the experimental spectrum, which provides more evidence that the molecule sampled in the experiments is a triplet.

\begin{figure}
\includegraphics[scale=0.5]{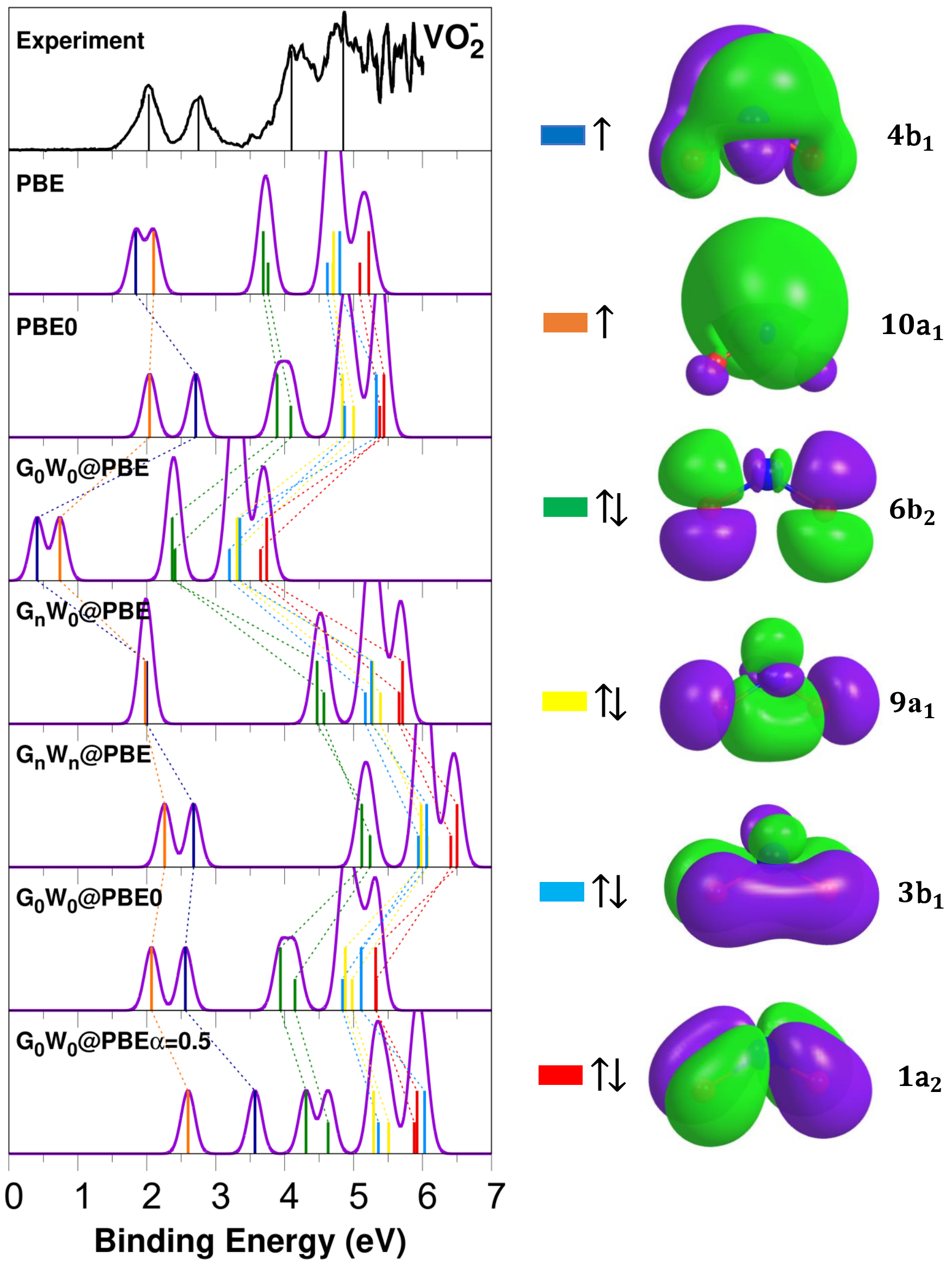}
\caption{\label{fig:VO2} Experimental photoelectron spectrum of VO$_{2}^-$ (Ref.~\onlinecite{Wu1998-b}) along with spectra computed with shifted PBE, shifted  PBE0, $G_0W_0@\text{PBE}$, $G_nW_0@\text{PBE}$, $G_nW_n@\text{PBE}$, $G_0W_0@\text{PBE0}$ and $G_0W_0$@PBE{\scriptsize$\alpha=0.5$}. Contour plots for some of the occupied molecular orbitals, in both majority ($\uparrow$) and minority ($\downarrow$) spin channels, are shown on the right, with matching color codes displayed in the spectra. For doubly occupied orbitals, the calculated energy displayed with the higher (lower) height refers to the orbital in the majority (minority) spin channel.}
\end{figure}

Figure \ref{fig:VO2} shows the experimental photoelectron spectrum of VO$_2^-$ obtained with a 193 nm laser\cite{Wu1998-b} and our spectra calculated at various levels of theory. The experimental spectrum consists of four main peaks at 2.03, 2.75, 4.10, and 4.85 eV. In our comparisons with the third and fourth experimental peaks, we take the average of our calculated spin-up and spin-down eigenvalues of doubly occupied orbitals. The vertical IPs calculated with PBE and PBE0 are 1.84 and 2.04 eV, the latter being in nearly perfect agreement with experiment. The other most obvious difference between shifted PBE and PBE0 spectra is the splitting of the first two peaks, which are calculated as 0.26 and 0.67 eV, respectively, compared to the experimental value of 0.72 eV. In fact, the whole shifted PBE0 spectrum has excellent agreement with experimental data with an MAE of 0.06 eV. The agreement with shifted PBE, on the other hand, is not as good with a MAE of 0.34 eV.

The predictions from the one-shot $GW$ approximation with PBE, PBE0, and PBE{\scriptsize$\alpha=0.5$} starting points follow the similar trends discussed earlier. $G_0W_0$@PBE predictions are very poor, while $G_0W_0$@PBE0 provides excellent agreement with experiment with an MAE of 0.09 eV, very similar to shifted PBE0. More exact exchange in the starting point worsens the agreement with experiment by overbinding all orbitals significantly, leading to an MAE of 0.58 eV at the $G_0W_0$@PBE{\scriptsize$\alpha=0.5$} level of theory. Eigenvalue self-consistency in $G$ with PBE starting point interestingly does not provide as good agreement with experiment compared to other molecules. While the $G_nW_0$@PBE prediction for the IP is very good (1.98 eV), the second peak predicted at 2.00 eV is underestimated by 0.75 eV. Here, we should also note that the HOMO eigenvalue calculated with $G_nW_0$@PBE does {\em not} correspond to the $4b_1$ orbital (which is the HOMO at PBE and $G_0W_0$@PBE levels), but to the $10a_1$ orbital. In particular, even though $G_nW_0$ predicts the quasiparticle energies of the $10a_1$ orbital to be lower than those of the $4b_1$ orbital calculated with aug-cc-pVDZ, aug-cc-pVTZ, and aug-cc-pVQZ basis sets, the eigenvalue difference gets smaller as the basis set size increases, such that when the eigenvalues are extrapolated, the $10a_1$ orbital becomes the HOMO. This difference in the convergence behavior of $10a_1$ and $4b_1$ orbitals is related to the fact that the $4b_1$ orbital has a much higher $3d$ content, and increasing the basis set size for such localized orbitals typically leads to a much larger lowering of the corresponding quasiparticle energies.\cite{Hung2017a} At the $G_nW_n$@PBE level, the predicted splitting between the $10a_1$ (now clearly the HOMO) and $4b_1$ (HOMO-1) orbitals significantly improves, however, the peaks with higher BE are overbound at this level of theory, leading to a MAE of 0.64 eV.

%%%%%%%%%%%%%%%%%%%%%%%%%%%%%%%%%%%  CrO2-1 %%%%%%%%%%%%%%%%%%%%%%%%%%%%%%%%%%

\subsection{\label{sec:CrO2-} CrO$_\textbf{2}^-$}

\begin{figure}
\includegraphics[scale=0.5]{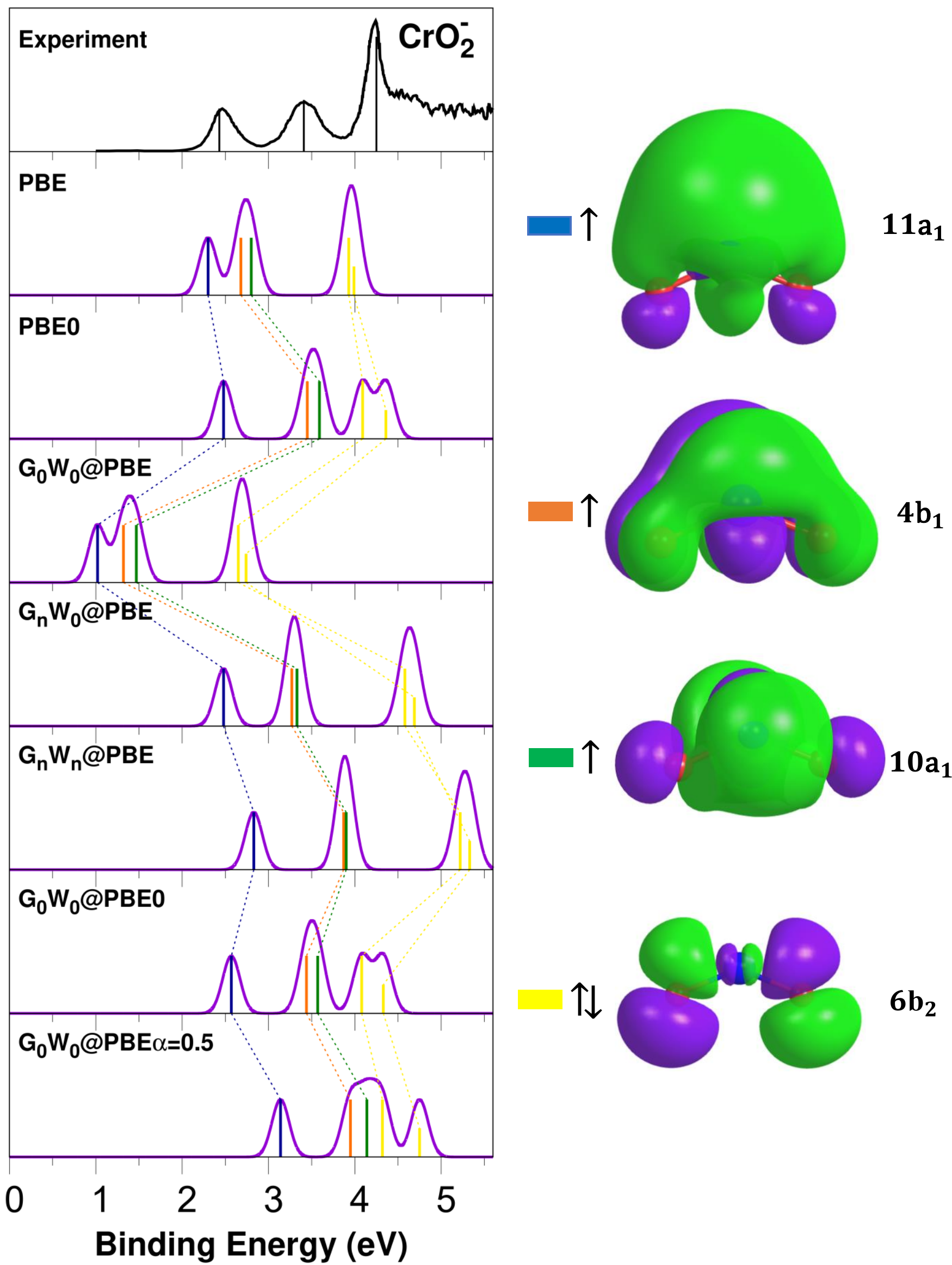}
\caption{\label{fig:CrO2} Experimental photoelectron spectrum of CrO$_{2}^-$ (Ref.~\onlinecite{Gutsev2001}) along with spectra computed with PBE, PBE0, $G_0W_0@\text{PBE}$, $G_nW_0@\text{PBE}$, $G_nW_n@\text{PBE}$, $G_0W_0@\text{PBE0}$ and $G_0W_0$@PBE{\scriptsize$\alpha=0.5$}. Contour plots for some of the occupied molecular orbitals, in both majority ($\uparrow$) and minority ($\downarrow$) spin channels, are shown on the right, with matching color codes displayed in the spectra. For doubly occupied $6b_2$ orbital, the calculated energy displayed with the higher (lower) height refers to the orbital in the majority (minority) spin channel.}
\end{figure}

Figure \ref{fig:CrO2} shows the experimental photoelectron spectrum of CrO$_2^-$ obtained with a 193 nm laser\cite{Gutsev2001} along with the computed spectra. There are three main peaks in the experimental spectrum at energies 2.43, 3.41, and 4.25 eV. With both PBE and PBE0 functionals, we find the ground state of CrO$_2^-$ to be the $^4B_1$ quartet with the valence configuration $(6b_2)^2(10a_1)^1(4b_1)^1(11a_1)^1$. The ground state of the neutral molecule is the $^3B_1$ triplet with the electron removed from the $11a_1$ orbital. Unlike the HOMO of VO$_2^-$, which is a localized orbital with a large $3d_{xz}$ character, the HOMO ($11a_1$) of CrO$_2^-$ is a delocalized orbital of primarily Ti $4s$ character with some Ti $p_z$ and $3d_{x^2-z^2}$ admixture, similar to the $10a_1$ orbitals in TiO$_2^-$ and VO$_2^-$. The frontier orbitals of large ($\sim70$\%) $3d$ character in CrO$_2^-$ are HOMO-1 and HOMO-2, whose DFT eigenvalues undergo a large downward shift ($\sim$1.8 eV) in going from PBE to PBE0 compared to the HOMO eigenvalue, which undergoes a shift of $\sim$1.2 eV. As a result, $11a_1$ orbital remains as the HOMO with PBE0 also, and there is no orbital reordering as observed in VO$_2^-$. We also note that in the computed photoelectron spectra, since the Kohn-Sham eigenvalues or the $GW$ quasiparticle energies of the $10a_1$ and $4b_1$ states are quite close to each other, we assign their average to the measured peak at 3.41 eV (a similar assignment was done in Ref. \onlinecite{Gutsev2001}), and the measured peak at 4.25 is compared with the average of the spin-up and spin-down $6b_2$ orbital energies.

The vertical IPs calculated with PBE and PBE0 functionals are 2.30 and 2.48 eV, respectively, which are in reasonably good agreement with experiment. However, as shown in Fig. \ref{fig:CrO2}, the shifted PBE does not have a very good agreement with experiment overall (MAE of 0.36 eV), since the second peak is underestimated significantly. This is not surprising, as this peak is due to $10a_1$ and $4b_1$ orbitals with large $3d$ character, which both suffer from SIE. As expected, the addition of a fraction of exact exchange mitigates this problem, and the shifted PBE0 spectrum has a much better agreement with experiment with a very low MAE of 0.06 eV. The $GW$ predictions are overall similar to the general trends discussed earlier. One-shot $GW$ with PBE0 starting point has the best overall agreement with experiment with an MAE of 0.09 eV, and each predicted peak matches almost perfectly with the shifted PBE eigenvalues. Different from VO$_2^-$ (but in line with other molecules), the $G_nW_0$@PBE predictions are quite good with an MAE of 0.18 eV. As before, adding self-consistency in $W$ eigenvalues or one-shot $GW$ with a larger exact exchange fraction worsen the agreement with experiment by overbinding the orbitals, and $G_0W_0$@PBE predicts orbitals which are too underbound, resulting in a very poor MAE of 1.66 eV. 

%%%%%%%%%%%%%%%%%%%%%%%%%%%%%%%%%%%  MnO2-1 %%%%%%%%%%%%%%%%%%%%%%%%%%%%%%%%%%

\subsection{\label{sec:MnO2-} MnO$_\textbf{2}^-$}

\begin{figure}
\includegraphics[scale=0.5]{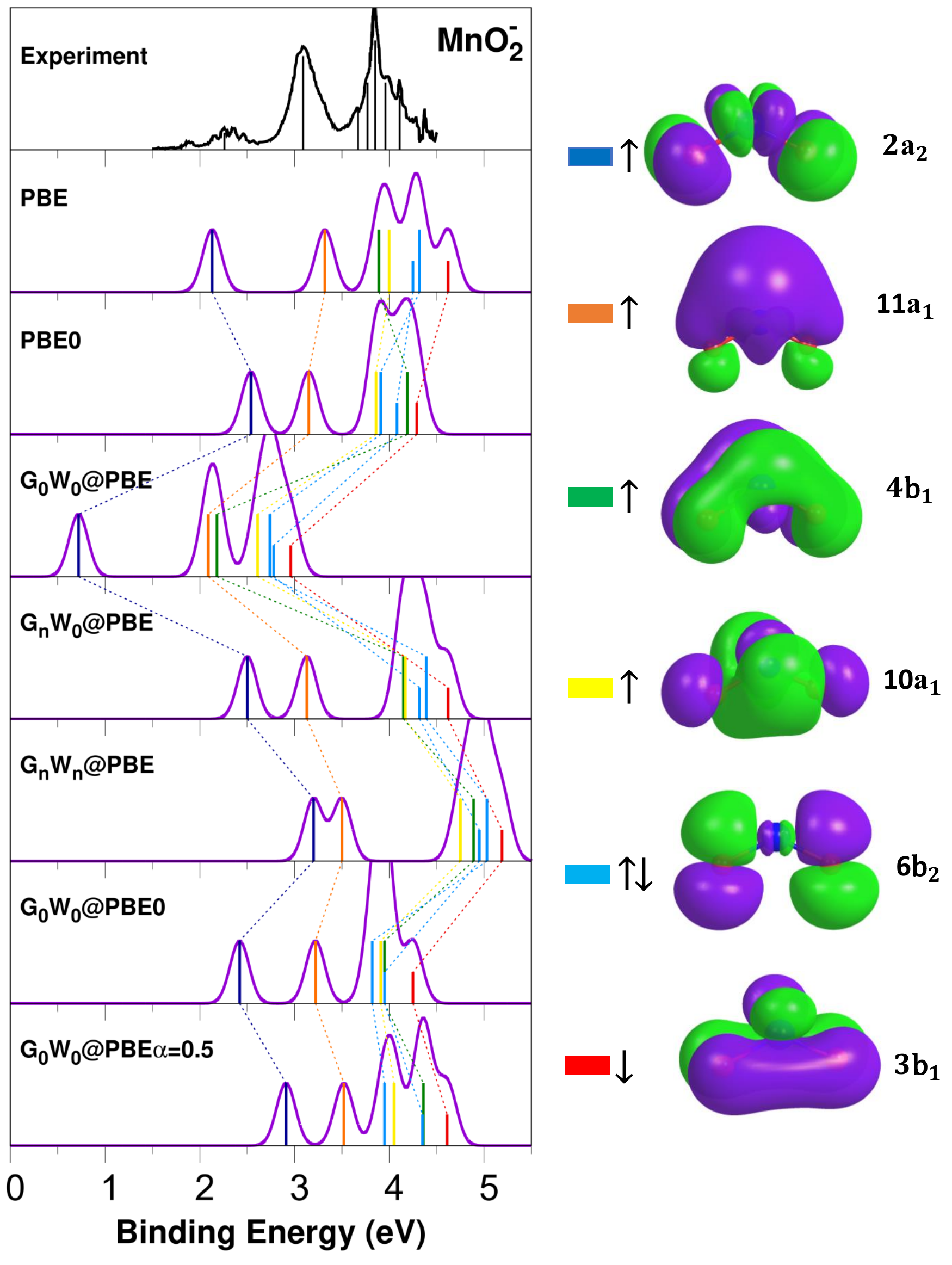}
\caption{\label{fig:MnO2} Experimental photoelectron spectrum of MnO$_{2}^-$ (Ref.~\onlinecite{Gutsev2000-b}) along with spectra computed with PBE, PBE0, $G_0W_0@\text{PBE}$, $G_nW_0@\text{PBE}$, $G_nW_n@\text{PBE}$, $G_0W_0@\text{PBE0}$ and $G_0W_0$@PBE{\scriptsize$\alpha=0.5$}. Contour plots for some of the occupied molecular orbitals, in both majority ($\uparrow$) and minority ($\downarrow$) spin channels, are shown on the right, with matching color codes displayed in the spectra. For doubly occupied $6b_2$ orbital, the calculated energy displayed with the higher (lower) height refers to the orbital in the majority (minority) spin channel.}
\end{figure}

In their combined experimental and theoretical study of the photoelectron spectra of MnO$_x^-$ clusters,\cite{Gutsev2000-b} Gutsev {\em et al.} found a strong temperature dependence of the spectrum for MnO$_2^-$. The experimental data displayed in Fig. \ref{fig:MnO2} shows the spectrum obtained with a 266 nm laser under cold conditions. The small peak near 1.89 eV was attributed to a higher energy isomer and will not be considered in the following discussion. The vertical IP of the MnO$_2^-$ was found to be 2.26 eV based on the analysis of the vibrational progression between 2.0 to 2.7 eV, followed by a broad peak at 3.09 eV. This is followed by a congested spectrum with many overlapping features between 3.6 and 4.3 eV, from which we have identified peaks at 3.67, 3.77, 3.85, 3.96, and 4.11 eV, giving a total of 7 peaks for comparison with our calculated spectra. 

With both PBE and PBE0 functionals, we find the ground state of MnO$_2^-$ to be the $^5B_2$ quintet with a valence configuration $(3b_1)^2(6b_2)^2(10a_1)^1(4b_1)^1(11a_1)^1(2a_2)^1$. We find the ground state of the neutral molecule to be the $^4B_1$ quartet with the electron removed from the $2a_2$ orbital, which is a localized molecular orbital derived primarily from the Mn $3d_{xy}$ atomic orbital hybridized with O $2p_x$ orbitals. Similar to the case of CrO$_2^-$, $11a_1$ is rather delocalized with significant $4s$ character, while $4b_1$ and $10a_1$ are both localized and derived primarily from Mn $3d_{xz}$ and $3d_{x^2-z^2}$, respectively. In the following discussion, we compare the energies of five majority-spin orbital energies and two minority-spin orbital ($6b_{2\downarrow}$ and $3b_{1\downarrow}$) energies with the seven peaks from the experimental spectrum.

The vertical IPs calculated with PBE and PBE0 functionals are 2.13 and 2.54 eV, respectively, compared to the experimental value of 2.26 eV. The deviation of the IP calculated with PBE0 from the experimental value is relatively large in comparison to other molecules considered here, however, the rest of the shifted PBE0 spectrum is in very good agreement with experiment leading to a MAE of 0.19 eV. While the IP calculated with PBE is very close to the experimental value, overall the shifted PBE spectrum with a MAE of 0.30 eV does not have as good agreement with experiment as PBE0. The most obvious difference between the two spectra is the splitting between the first two peaks, which are found to be 1.19 and 0.61 eV with PBE and PBE0, respectively, compared to the experimental value of 0.83 eV. This can be understood in terms of the difference in the spatial extents of the relevant orbitals, $2a_2$ and $11a_1$, and the downward shift in energy they experience with the addition of a fraction of exact exchange, as discussed for several cases earlier. In the case of MnO$_2^-$, the DFT eigenvalue of the more localized $2a_2$ orbital decreases by 1.87 eV, while that of the less localized $11a_1$ orbital decreases by 1.29 eV in going from PBE to PBE0 (keep in mind that the PBE and PBE0 spectra plotted in Fig. \ref{fig:MnO2} are {\em shifted} spectra).

The predictions from the variants of the $GW$ approximation are consistent with the overall trends discussed so far. Best agreement is achieved with $G_0W_0$@PBE0 (MAE of 0.12 eV), followed by $G_nW_0$@PBE with a MAE of 0.37 eV. While the first two peaks are predicted well with $G_nW_0$, states with higher BEs are overbound by $0.4-0.5$ eV. Adding eigenvalue self-consistency in $W$ overbinds these orbitals more, and MAE increases to 0.97 eV. PBE starting point for one-shot $GW$ again leads to the worst agreement with experiment (MAE of 1.23 eV), and $G_0W_0$@PBE{\scriptsize$\alpha=0.5$} overbinds all states by $0.3-0.6$ eV with a MAE of 0.43 eV. 

%%%%%%%%%%%%%%%%%%%%%%%%%%%%%%%%%%%  Trends among the clusters %%%%%%%%%%%%%%%%%%%%%%%%%%%%%%%%%%
\section{\label{sec:Trend}Discussion}

%%%%%%%%%%%%%%%%%%%%%%%%%%% Graph: IP and MAE  %%%%%%%%%%%%%%%%%%%%%%%%%%%%%%%%%%%%
%\comment{
\begin{figure*}
%\hfill
\includegraphics[scale=0.65]{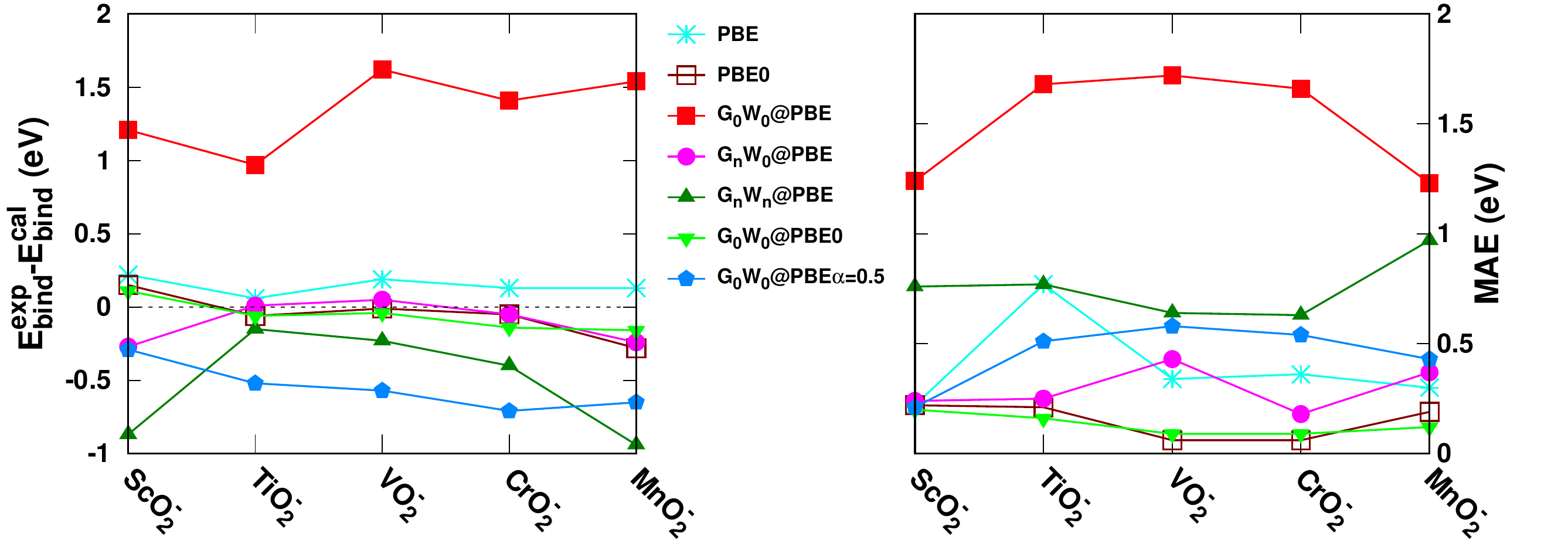}

\caption{\label{fig:MAE-IP} Difference between experimental and our calculated vertical IPs ($E_{\text{bind}}^{\text{exp}}-E_{\text{bind}}^{\text{cal}}$) (left), and mean absolute error (MAE) of calculated electron binding energies (right) of TMO$_2^-$ molecules at different levels of theory.}
\end{figure*}
%}

The overall trends for how the predictions from various levels of DFT and $GW$ theories compare with experimental data for all TMO$_2$ anions are summarized in Fig.~\ref{fig:MAE-IP}, which shows the differences between the computed and experimental vertical IPs (left panel) as well as the MAEs (right panel). We notice that the vertical IPs computed with the PBE functional are quite good, underestimated slightly ($0.1-0.2$ eV) with respect to experiment. However, when higher BE peaks are taken into account, the agreement with experiment worsens, particularly for TiO$_2^-$ with an MAE of 0.77 eV, which highlights the importance of comparing not just the vertical IP but all the higher BE peaks in available PES in assessing the performance of a particular level of theory. For shifted PBE0, on the other hand, not only are the predictions for vertical IPs in reasonably good agreement with experiment (typically less than 0.15 eV, but with an outlier for the case of MnO$_2^-$ where the deviation from experiment is 0.28 eV), but also the overall MAEs (less than $\sim0.2$ eV) are quite satisfactory.

For the case of one-shot $G_0W_0$@PBE{\scriptsize$\alpha$} level of theory, the starting point makes a striking difference. As highlighted for all TMO$_2$ anions in the previous section, a PBE starting point leads to very poor agreement with experiment for both the vertical IPs (average underestimate compared to experiment for all cases considered being 1.36 eV) and the MAEs, which are slightly higher. PBE0 starting point, on the other hand, leads to the best agreement overall with experimental data among all levels of theory considered in this study, not only for vertical IPs (within $\sim0.1$ eV), but also for higher BE peaks (with a similar MAE of $0.1-0.2$ eV). With increasing amount of exact exchange in the DFT starting point, however, the agreement with experiment deteriorates: With $G_0W_0$@PBE{\scriptsize$\alpha=0.5$}, all orbitals are overbound by $\sim 0.5$ eV, with the exception of ScO$_2^-$ where MAE remains near 0.2 eV, close to the value obtained with $\alpha=0.25$. This is consistent with the results obtained for ScO$^-$ in Ref. \onlinecite{Byun2019}, where any value of $\alpha$ in the DFT starting point in the range $0.25\le\alpha\le 1$ leads to good agreement with experimental data and does not shift the predicted quasiparticle energies significantly for this molecule, which was attributed to the lack of significant $3d$ character in the relevant orbitals. We also note that our finding of nearly excellent agreement of $G_0W_0$@PBE{\scriptsize$\alpha$} at the ``sweet spot'' value of $\alpha=0.25$ should not be taken as a general trend for the $G_0W_0$@PBE0 level of theory for TMO molecular systems, as this work has focused particularly on early TM dioxide systems. Typically, the amount of $\alpha$ needed for good agreement with experimental data in the DFT starting point of one-shot $GW$ calculations for TMO molecules is dependent on the amount of the $3d$ character of the orbitals\cite{Byun2019} as well as the content of TM in the molecule. For example, Shi {\em et al.}\cite{Shi2018} showed in their study of small copper oxide molecular anions that while $\alpha=0.5$ worked well for molecules like Cu$_2$O$^-$ and CuO$^-$, values near $\alpha=0.25$ were optimal for molecules with less Cu content, such as CuO$_2^-$ and CuO$_3^-$.

For the case of ev$GW$, updating eigenvalues only in $G$ ($G_nW_0$) with a PBE starting point leads to very good agreement with experiment for the vertical IPs, typically within 0.05 eV, except for the two slight outliers of ScO$_2^-$ and MnO$_2^-$ where the deviation is $\sim 0.25$ eV. When higher BE peaks are considered, the agreement with experiment slightly worsens, but the MAE value of 0.3 eV averaged over the five TMO$_2$ anions is still reasonable. Adding self-consistency in $W$, on the other hand, overbinds all orbitals, as the poles of the self-energy become more negative (for occupied orbitals) with the increase in the neutral excitation energies. For vertical IPs of TiO$_2^-$, VO$_2^-$, and CrO$_2^-$, $G_nW_n$@PBE predictions are in good agreement with experiment (overestimated by 0.11, 0.23, and 0.40 eV, respectively), but the predicted IPs for ScO$_2^-$ and MnO$_2^-$ are not, as they are overestimated by $\sim 0.9$ eV. When all peaks are considered, $G_nW_n$@PBE is clearly not a predictive level of theory, as the MAE averaged over the five TMO$_2$ anions is $\sim$ 0.75 eV.

\begin{figure}
\includegraphics[scale=0.32]{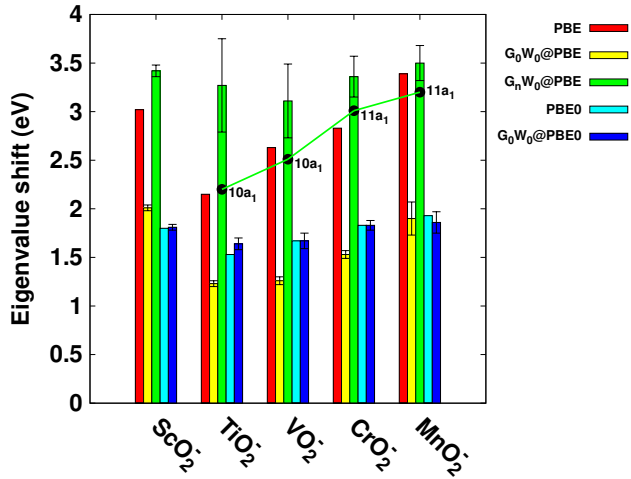}
\caption{\label{fig:plot_shift_DFT_GW} Calculated eigenvalue shifts for TMO$_2^-$ molecules at each level of theory. For PBE and PBE0, the bars show the magnitudes of shifts needed to align the HOMO eigenvalues with negative of the computed vertical IPs. For $GW$ levels of theory, the bars show the magnitudes of the shifts of the $GW$ eigenvalues from the eigenvalues of the corresponding DFT starting points averaged over the measured peaks. The error bars are the standard deviations of the shifts for each level of $GW$ theory. The black dots show the shifts for the delocalized $10a_1$ and $11a_1$ orbitals discussed in the text. This orbital is unoccupied in ScO$_2^-$.}
\end{figure}

In order to gain more insight into the successes and failures of various levels of $GW$ theories for the TMO$_2$ anions considered, we now focus on the (averaged) eigenvalue shifts for each molecule, as shown in Fig. \ref{fig:plot_shift_DFT_GW}. For the $GW$ levels of theory, the bars show for each TMO$_2$ anion the magnitudes of the shifts of the $GW$ eigenvalues from the corresponding DFT starting point eigenvalues averaged over the measured peaks. The standard deviations (SDs) of the shifts are shown with error bars for each level of $GW$ theory. For PBE and PBE0, the bars show the magnitudes of the shifts needed to align the HOMO eigenvalue with negative of the computed vertical IP of the anion discussed earlier. One immediate observation from Fig. \ref{fig:plot_shift_DFT_GW} is that the eigenvalue shifts for the $G_0W_0$@PBE0 (blue bars) are very close to the PBE0 shifts (cyan bars) for all molecules considered. This, combined with the observation that the SDs of the shifts for $G_0W_0$@PBE0 are quite small (less than 0.1 eV), shows that the $G_0W_0$@PBE0 spectra can be obtained by almost a uniform shift of the PBE0 spectra by the amount needed to align HOMO with negative the vertical IP of the anion, irrespective of particular nature (localized, delocalized, large or small $3d$ character) of the orbitals, resulting also in similar MAEs with respect to experimental data. The very good agreement of the $G_0W_0$@PBE0 level of theory with experimental PES can then be attributed to the correct energetic positioning of the orbitals obtained with PBE0 and the PBE0 wave functions being good approximations to the true quasiparticle wave functions, so that a simple first-order $G_0W_0$ perturbative correction is enough to lead to accurate quasiparticle energies. 

%%%%%%%%%%%%%%%%%%%%%%%%%%% Graph: Eigenvalue shift  %%%%%%%%%%%%%%%%%%%%%%%%%%%%%%

With a PBE starting point, on the other hand, the average shifts for $G_0W_0$@PBE (yellow bars in Fig. \ref{fig:plot_shift_DFT_GW}) are much smaller than the PBE eigenvalue shifts required to align HOMO with negative the vertical IP (red bars in Fig. \ref{fig:plot_shift_DFT_GW}). That is, unlike the very good agreement of shifted PBE0 and $G_0W_0$@PBE0 with each other (as well as with experiment), application of first-order $G_0W_0$ perturbative correction to the PBE eigenvalues using PBE wave functions fails to provide the required amount of downward shift to achieve a reasonably good level of agreement with experimental data. To quantify this further, we compared the PBE and PBE0 eigenvalues for all TMO$_2$ anions and considered the amounts of shifts that would be needed to achieve perfect agreement with experimental data. Averaged over all experimental peaks and all TMO$_2$ anions, the required shift is $3.09\pm 0.27$ eV and $1.79\pm0.13$ eV for PBE and PBE0, respectively, showing (i) how underbound PBE orbitals are for perturbative corrections to be effective, and that (ii) the required shifts for PBE have a much larger variation than those for PBE0 due to the significant difference in the SDs.

These last two observations are also consistent with the trends of eigenvalue shifts for $G_nW_0$@PBE shown with green bars in Fig. \ref{fig:plot_shift_DFT_GW}. In this case, the shifts are significantly larger, correcting the eigenvalues of the underbound PBE orbitals, and the amount of shift depends on the nature of the particular orbital, due to the large SD observed for most TMO$_2$ molecular anions, with the exception of ScO$_2^-$, where the occupied frontier orbitals are composed of primarily O $2p$ states and have similar spatial extents. The large SD observed in $G_nW_0$@PBE eigenvalue shifts is primarily due to a particular orbital with $a_1$ character ($10a_1$ in TiO$_2^-$ and VO$_2^-$, and $11a_1$ in CrO$_2^-$ and MnO$_2^-$), which is either the HOMO or HOMO-1 in the majority spin channel depending on the level of theory (see Fig. \ref{fig:VO2} for VO$_2^-$) or the molecule (in MnO$_2^-$, HOMO is the $2a_2$ orbital). The shifts for this particular orbital also displayed in Fig. \ref{fig:plot_shift_DFT_GW} are significantly less than the averaged shift, {\em e.g.} the $G_nW_0$@PBE eigenvalue for the $10a_1$ state of TiO$_2^-$ is 2.2 eV lower than the PBE eigenvalue, which is much smaller than the average downward shift of 3.27 eV. As a result, the SD is particularly high at the $G_nW_0$@PBE level of theory, as different orbitals are shifted by varying amounts to achieve a reasonably good level of agreement with experiment. As mentioned earlier, this particular $a_1$ orbital has a large TM $4s$ character with some TM $p_z$ and $3d_{x^2-z^2}$ admixtures. It is important to note that while the $3d$ character is not very large, it is still appreciable, ranging from 15 to 30\% in going from TiO$_2^-$ to MnO$_2^-$. The primary reason why this orbital does not undergo a large downward shift is its very delocalized nature in spite of having an appreciable $3d$ content, consistent with the observation that $G_nW_0$@PBE does a particularly good job for the vertical IPs, and the agreement deteriorates slightly for the more localized higher BE peaks. 

%%%%%%%%%%%%%%%%%%%%%%%%%%%%%%%%%%%  Summary %%%%%%%%%%%%%%%%%%%%%%%%%%%%%%%%%%
\section{\label{sec:summary}Summary}

In summary, we provided a detailed comparison of the DFT- and $GW$-computed eigenvalue spectra to experimental photoelectron spectra of five $3d$-transition metal dioxide molecular anions, ScO$_2^-$, TiO$_2^-$, VO$_2^-$, CrO$_2^-$, and MnO$_2^-$. We focused our study on the comparison between semi-local and hybrid functional predictions (within DFT and as a starting point for $G_0W_0$ calculations) as well as the effects of eigenvalue self-consistency with a semi-local DFT starting point. Overall, both the shifted PBE0 and $G_0W_0$@PBE0 appear to stand out as the best levels of theory, as they are able to reproduce the experimental BEs with $0.1-0.2$ eV accuracy (for some peaks, especially the vertical IPs, with much better accuracy). With shifted PBE, on the other hand, while the total energy differences between the anion and the neutral molecule at the geometry of the anion lead to similarly accurate vertical IPs, the agreement with experiment quickly deteriorates for higher BE peaks. A one-shot $GW$ calculation with a semi-local starting point like PBE is clearly a poor choice leading to the worst agreement with experiment (average error $\sim$1.5 eV or more), as the PBE orbitals are too underbound and the PBE wave functions are too delocalized for a perturbative self-energy correction to be effective. These observations highlight the importance of adding a fraction of Fock exchange to mitigate self-interaction errors in predicting the quasiparticle energies of transition metal oxide molecular systems within DFT or $GW$ levels of theory. However, too much Fock exchange (e.g. $\alpha=0.5$) also worsens the agreement with experiment by overbinding orbitals. We found that $G-$only eigenvalue self-consistent $GW$ with $W$ computed from PBE (G$_nW_0$@PBE), while not as accurate as shifted PBE0 or $G_0W_0$@PBE0, still provides a reasonably good description of the quasiparticle energies for occupied frontier orbitals in these transition metal oxide molecular systems. Updating the eigenvalues in $W$ ($G_nW_n$@PBE), on the other hand, leads to poor agreement with experiment overall, even though it may be reasonably accurate for vertical IPs of some of the molecular anions considered. We caution the reader that our finding of very good agreement of $G_0W_0$@PBE0 predictions with experimental data for the particular case of these early $3d-$transition metal dioxide molecular anions should be viewed in the context of our previous studies on other transition metal oxide molecular systems,\cite{Shi2018,Byun2019} which showed that higher values of Fock exchange might be needed in the DFT starting point of $GW$ calculations to achieve good agreement with experiment in systems with more localized orbitals and higher transition metal content. When viewed from this perspective, the results presented in this paper still lead us to advocate using $G_nW_0$@PBE as a practical $GW$ scheme that presents a good balance between accuracy and efficiency in predicting quasiparticle energies of transition metal oxide molecular systems without the need for a system-dependent parameter in the starting DFT description. Future studies on the more challenging late $3d-$transition metal dioxide molecular anions, such as FeO$_2^-$, NiO$_2^-$ and CuO$_2^-$, and other transition metal oxide molecules with varying amounts of transition metal content are needed to further assess and understand the predictive capabilities of $GW$ methods for electronic excitations in molecular systems with moderate electron correlation.

\section*{supplementary material}

See \textcolor{blue}{supplementary material} for more details, results, and discussion.

\section*{acknowledgments}

This work was supported by the U.S. Department of Energy Grant No. DE-SC0017824. This research used resources of the National Energy Research Scientific Computing Center, a DOE Office of Science User Facility supported by the Office of Science of the U.S. Department of Energy under Contract No. DE-AC02-05CH11231. We would also like to thank Young-Moo Byun for useful discussions in the earlier stages of this work.

\section*{references}
\nocite{*}

\bibliography{Sc-Ti-V-Cr-Mn_TMO2-}

\end{document}